\documentclass[aps,prl,reprint,superscriptaddress]{revtex4-2}

\usepackage{graphicx}
\usepackage{dcolumn}
\usepackage{bm}
\usepackage{tabularx}
\usepackage{siunitx}
\usepackage{xcolor}
\usepackage{comment} 
\usepackage{amsmath} 
\DeclareSIUnit{\sps}{Sps}

\begin{document}

\title{Temporal limitations and digital data processing in continuous variable measurements of non-Gaussian states}

\author{Antoine Petitjean}
\affiliation{Universit\'e C\^ote d'Azur, CNRS, Institut de Physique de Nice, 17 av. J. Laupretre, 06200 Nice, France}

\author{Anthony Martin}
\affiliation{Universit\'e C\^ote d'Azur, CNRS, Institut de Physique de Nice, 17 av. J. Laupretre, 06200 Nice, France}

\author{Mohamed F. Melalkia}
\affiliation{Universit\'e C\^ote d'Azur, CNRS, Institut de Physique de Nice, 17 av. J. Laupretre, 06200 Nice, France}

\author{Tecla Gabbrielli}
\affiliation{European Laboratory for Non-linear Spectroscopy (LENS), University of Florence, Via Nello Carrara 1, 50019 Sesto Fiorentino (FI), Italy}
\affiliation{Istituto Nazionale di Ottica (CNR-INO), CNR, Largo Enrico Fermi 6, 50125 Firenze, Italy}

\author{L\'eandre Brunel}
\affiliation{Universit\'e C\^ote d'Azur, CNRS, Institut de Physique de Nice, 17 av. J. Laupretre, 06200 Nice, France}

\author{Alessandro Zavatta}
\affiliation{European Laboratory for Non-linear Spectroscopy (LENS), University of Florence, Via Nello Carrara 1, 50019 Sesto Fiorentino (FI), Italy}
\affiliation{Istituto Nazionale di Ottica (CNR-INO), CNR, Largo Enrico Fermi 6, 50125 Firenze, Italy}

\author{S\'ebastien Tanzilli}
\affiliation{Universit\'e C\^ote d'Azur, CNRS, Institut de Physique de Nice, 17 av. J. Laupretre, 06200 Nice, France}

\author{Jean Etesse}
\affiliation{Universit\'e C\^ote d'Azur, CNRS, Institut de Physique de Nice, 17 av. J. Laupretre, 06200 Nice, France}

\author{Virginia D'Auria}
\email{virginia.dauria@univ-cotedazur.fr}
\affiliation{Universit\'e C\^ote d'Azur, CNRS, Institut de Physique de Nice, 17 av. J. Laupretre, 06200 Nice, France}
 %% email address is required; see note below about the corresponding author designation

% use {asbstract*} to suppress the copyright line. Copyright information will be added in production

\begin{abstract} 
Non-Gaussian quantum states and operations are essential tools for multiple quantum information protocols exploiting light as information career. In this context, a key role is played by schemes operating with continuous wave light, in which non-Gaussian states are obtained by photon subtraction/addition and eventually reconstructed by quantum state tomography. In these configurations, the temporal resolution of the homodyne detection and the digital data processing critically affect our ability to faithfully reconstruct the produced non-Gaussian states. In this work, we apply digital data processing to experimental data to study how the temporal performances of the detection chain affect the acquisition and treatment of tomographic data. This allows understanding how these features impact  the quality of quantum states observed by non-ideal detection chains. By doing so, we discuss the actual constraints on the acquisition and reconstruction of non-Gaussian states by taking into account the limitations of realistic experimental resources.

\end{abstract}
\maketitle

%%%%%%%%%%%%%%%%%%%%%%%%%%  body  %%%%%%%%%%%%%%%%%%%%%%%%%%
\section*{Introduction}
 Non-Gaussian (NG) states of light are essential resources for quantum optics and its applications in quantum communication and computing exploiting a continuous variable (CV) encoding of quantum information~\cite{asavanant2022optical}. In experiments, they can be conveniently produced by performing a photon-count measurement on a subsystem of an entangled state: a given measurement result heralds the successful preparation of the desired state on the remaining modes of the initial entangled resource~\cite{lvovsky2020production, RaMultiNG2020}. In CV schemes, the field quadratures of heralded NG states are measured by homodyne detectors (HDs) triggered by a photon-count signal: the collection of quadrature data enables a complete state reconstruction via quantum tomography~\cite{Lvovsky_Tomo2009}. This protocol has been successfully used to detect single photon states~\cite{morin2012high, lvovsky2020production}, Schroedinger kitten states~\cite{ourjoumtsev2006quantum, melalkia_plug-and-play_2022, Takase:22}, hybrid entangled states~\cite{ZavattaHybrid2014, MorinHybrid2014}, multimode NG state~\cite{RaMultiNG2020} and, more recently, Gottesman-Kitaev-Preskill (GKP) optical states~\cite{FurusawaGKP2024, GKPXanadu2025}. 

In conditional state preparation, imperfections on the heralding detector (see~\cite{Gouzien_2018}) or on the setup used to measure the heralded state critically affect the capability of preparing and observing NG states. Optimal access to prepared NG states demands, in particular, the ability to extract from the set of HD data those corresponding to a specific mode with a temporal profile, $u_\text{id}(t)$~\cite{morin2012high, RaMultiNG2020, melalkia2022theoretical}. The actual shape of $u_\text{id}(t)$ is determined by both the frequency/time features of the entangled state source (ESS) used in protocol and the optical setup employed to perform the heralding operation~\cite{asavanant2017generation, RaMultiNG2020}. In the following we will refer to the mode associated to $u_\text{id}(t)$ as the ideal detection mode. 
 The scope of this work is to understand how temporal features of the CV detection chain drive us away from the best measure of the heralded quantum states. We focus on continuous wave (CW) experiments, where both the pump of the ESS and the local oscillator (LO) of the HD are monochromatic lasers. This configuration has been chosen for multiple ground-breaking CV demonstrations~\cite{MorinHybrid2014, FurusawaGKP2024, Sonoyama2025, lenzini2018integrated} because it simplifies the management of spurious non-linear effects, chromatic dispersion and optical mode-matching. In addition, the detection mode can be reconstructed directly from quadratures' data~\cite{morin2013experimentally, MorinHybrid2014, melalkia_plug-and-play_2022}. To comply with time degrees-of-freedom in CW schemes, CV measurement chains must be equipped with adequate electronics able to accurately resolving and digitalizing data in the temporal mode $u_\text{id}(t)$. The shape of $u_\text{id}(t)$ can be smoothened by using narrow-band optical filters on the heralding path but this comes at the price of drastically reduced heralding rates. To mitigate this trade-off, most previous experiments have relied on very fast, and often costly, continuous-variable detection chains, without a systematic analysis of the minimal temporal requirements necessary for reliable quantum state reconstruction. Our work follows a different approach. It aims to investigate how stringent actually are the conditions on the temporal performances of the CV acquisition system. The objective is to study the robustness of state detection and tomographic reconstruction against experimental limitations. We will focus our attention on the HD electrical bandwidth and the data sampling rate that are key parameters in temporal signals' acquisition. Reported results exploit experimental HD data that we obtained in the context of Schroedinger kittens' preparation by heralded single photon subtraction from a squeezed state~\cite{melalkia_plug-and-play_2022}. Such a choice allows us to perform our analysis on real HD photo-currents rather than on numerically simulated ones. The global scope of this work is to identify convenient and feasible experimental conditions that guarantee a faithful reconstruction of heralded NG states, while possibly relaxing hard restrictions on the detector's performances. While we do not aim to derive general criteria, we illustrate and rationalize nontrivial effects due to realistic limitations in detection, that have clear experimental implications and provides tools to lighten the design of future quantum-optics experiments. By virtue of its generality, our methodology and results can thus be extended to different scenarios, including those involving more complex NG operations~\cite{MorinHybrid2014, FurusawaGKP2024, GKPXanadu2025}. The impact of our investigation, however, goes well beyond this specific situation as it applies, in a general way, to any scheme in which temporal features are critical in quantum state reconstruction. As CV quantum optics progresses toward more complex and multiplexed architectures, understanding and controlling its practical limitations becomes increasingly important. In this regard, we note that the interest in temporal-mode effects extends beyond NG-state preparation and is highly relevant in other domains, including continuous-variable quantum key distribution~\cite{Liu2022}, optical frequency comb metrology~\cite{Lordi2024} and quantum state detection via dual homodyne measurement~\cite{Takase2019}.
\section{Heralded acquisition of CV experimental data}\label{sec:setup}

\begin{figure}[h]
    \centering
    \includegraphics[width=\columnwidth]{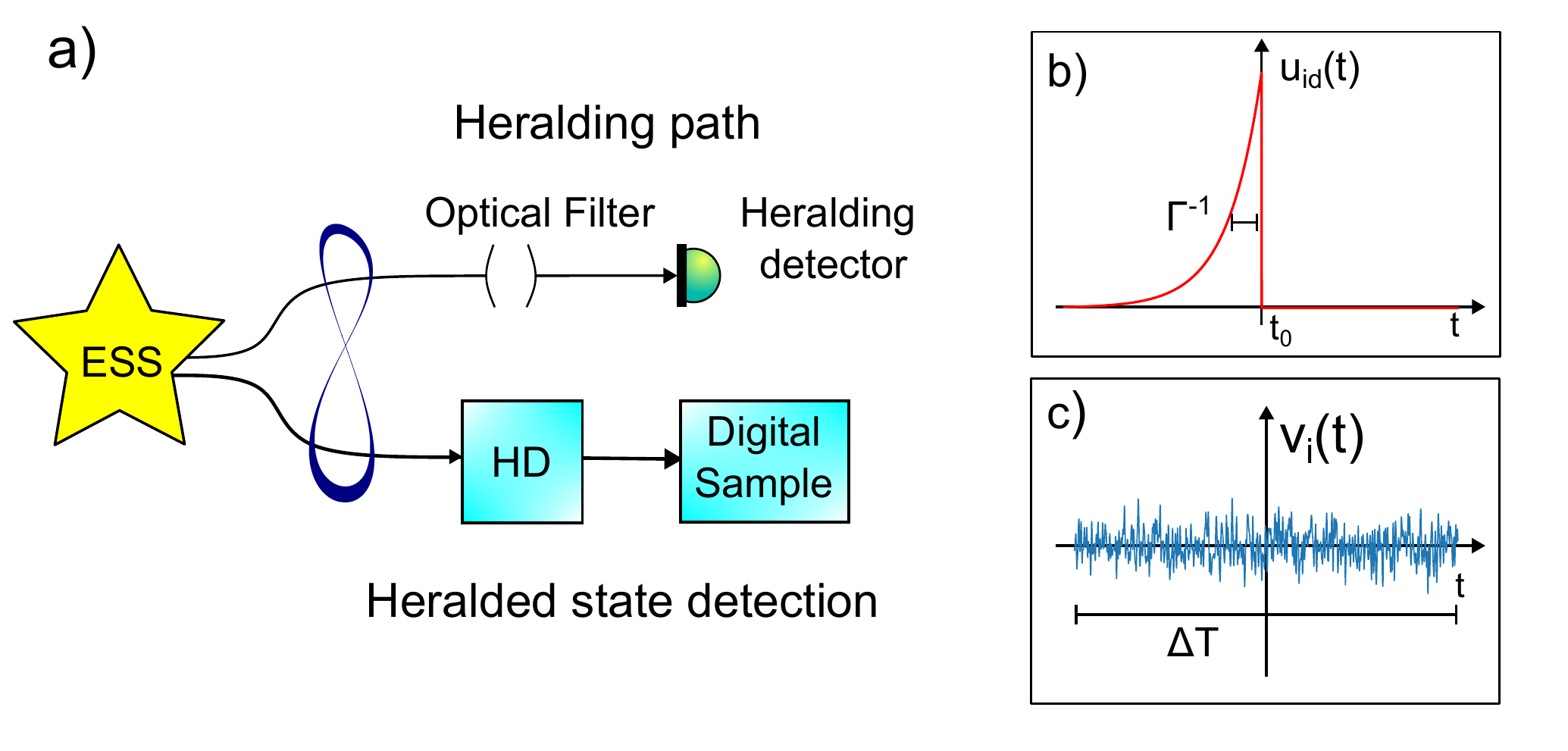}
    \caption{a) Schematic diagram of an heralding scheme. ESS stands for entangled state source. The heralding path contains a filtering stage and single photon detection. On the heralded state detection, the homodyne detection and a digital sampling part have finite bandwidths and sampling rate, respectively. b) Temporal profile of the ideal detection mode in the case of an optical cavity used as filter on the heralding path and of a single pass ESS (\textit{i.e.} a source without a feedback or resonator mechanism). c) Digitalized trace at the output of the CV detection chain that measures the heralded state. Each heralding event leads to the recording of one of such traces of duration $\Delta T$. The heralding path detector is taken to be ideal; the effect of non ideal heralding on the quality of produced NG states has been already treated in a previous paper~\cite{Gouzien_2018}.}
    \label{fig:enter-label}
\end{figure}
In all heralded protocols, the optical modes relevant to the experiment are the one in which the heralding detection takes place and the one occupied by the produced NG state, \textit{i.e.} $u_\text{id}(t)$: all other modes do not carry any relevant information and must be kept unobserved~\cite{Molmer2007}. This requires, on the heralded state side, the ability to perform a projective measurement on the sole mode $u_\text{id}(t)$, that, in turn, must be known or accessible to experimentalists. In an ideal case, for a fixed LO optical phase $\vartheta$, the HD yields a measure of the quadrature $\hat{X}(\vartheta)=\frac{1}{\sqrt{2}}(\hat{a}e^{-i\vartheta}+\hat{a}^\dagger e^{i\vartheta})$ where $\hat{a}$ and $\hat{a}^\dagger$ are the bosonic operators of light in the ideal detection mode. This operator can be written as $\hat{a}=\int u_\text{id}(t) \hat{a}(t) dt$, where $\hat{a}(t)$ are the bosonic operators associated to temporal modes corresponding to a well define time $t$. Realistic scenarios, however, can suffer from a poor determination of $u_\text{id}(t)$ and from HD data distortion due to inadequate signal acquisition. These combined effects blur the HD data relative to the target state and mix them with those carried by other optical modes that impinge on the detector (\emph{e.g.} vacuum or squeezed vacuum states). To understand these concepts, in this section, we will define the mode of interest in heralded state preparation and recall the different conceptual steps of CV data acquisition so as to underline the potential limitations to a correct state reconstruction in non-ideal situations.

\subsection{Mode definition}
Figure~\ref{fig:enter-label}-a shows a general architecture for the conditional preparation of NG states: two (or more) entangled optical systems from an ESS are spatially separated; the result of the detection of one sub-system, generally a photon count, heralds the preparation of the target quantum state on the other one(s). In practical situations, ESS are based on non-linear optical processes. Depending on the working point, these can give rise to multipartite correlations among different optical modes: these latter, if unmastered, can degrade the quality of heralded states~\cite{melalkia2022theoretical}. To reduce the impact of unwanted correlations, a very common strategy consists in inserting, before the heralding detector, an optical filter of line-width $\Gamma$ small compared with the ESS emission bandwidth~\cite{yoshikawa2017purification, melalkia2022theoretical}. 

The choice of the filter on the heralding path determines the actual shape of the mode $u_\text{id}(t)$ on the heralded state's side~\cite{Molmer2007}. Here, we will consider a cavity-based filter~\cite{yoshikawa2017purification, melalkia_plug-and-play_2022} and, without any loss of generality, the case of single pass ESS i.e. of a source based on a non linear interaction without any optical cavity or feedback mechanism, as available in many guided-wave experiments~\cite{asavanant2017generation, lenzini2018integrated}. These choices lead to $u_\text{id}(t)$ shaped as in Figure~\ref{fig:enter-label}-b. The left side of the function has an exponential behavior, whose full width at half maximum $\sim\Gamma^{-1}$ corresponds to the inverse of cavity spectral line-width, typically of the order of few tens of ns (\emph{i.e.} $\Gamma\sim$ 10 to 100\,MHz)~\cite{morin2012high, melalkia_plug-and-play_2022}. The sharp right side exhibits a rapid decrease whose characteristic time is of the order of the photons' coherence time at the ESS output~\cite{morin2012high, asavanant2017generation}: in single pass sources pumped in CW, this time is extremely small ($\sim$100\,fs) almost giving a step-like function. The profile asymmetry reflects the causality principle that forbids the heralded state to perdure after the instant $t_0$ at which the heralding detector clicks~\cite{yoshikawa2017purification}. 
In what follows, we will indicate as $B$ the spectral bandwidth of the ideal detection mode in the Fourier domain: this quantity indicates the frequency range over which the profile's spectral density is above -3\,dB of his maximum. Note that, due to the sharp edge of $u_\text{id}(t)$, the actual value of $B$ is in general wider than $\Gamma$.

 \subsection{Homodyne detection of heralded states}

Heralding events at times $t_{i}$ trigger the CV acquisition of the produced NG states. The HD detection relies on the detection of the optical interference obtained by sending (BS) the LO and the signal under scrutiny on a 50:50 beam-splitter. The output of the BS are send to two balanced photodiodes whose outputs are subtracted. This process converts the interference signal into a photocurrent, which is then amplified by dedicated electronic stages to produce an analog voltage signal that carries information about the state quadratures over time.

In CW experiments, the LO is a (quasi-)monochromatic laser while the signal beam generally contains multiple monochromatic spectral components. In these conditions, the HD photocurrent is made of multiple beat notes oscillating at frequencies $f$, that arise from the interference of the monochromatic LO with the individual optical components of the signal. Ideally, fast balanced detections should be able to acquire and to follow all beat components including those at very high $f$. In real systems, the “quantity” of signal components that are actually detected depend on the electronics bandwidth of the HD detector that thus play a key role. Realistic HD are, indeed, only able to deal with frequency components below a certain cutoff frequency $f_c$: given an input signal $s(t)$ with Fourier transform $\tilde{s}(f)=FT[s(t)]$, the detector output will be $s_\text{fc}(t)=FT^{-1}[H(f,f_c)\cdot\tilde{s}(f)]$, with $H(f,f_c)$ a filter function. 
 
 The response of most of reported HD circuits can be modeled as a $2^{nd}$ order Butterworth low-pass filter~\cite{kumar_versatile_2012, Tasker_2020}. In the Fourier transform domain, its associate gain function reads as:
\begin{equation}
    H(f,f_c)=\frac{G_0}{\sqrt{1+(f/f_c)^4}},
    \label{TransfertF}
\end{equation}
with $f_c$ the -3\,dB cutoff frequency and $G_0$ the low frequency gain (\textit{i.e.} at $f=0$). The actual value of $f_c$ is directly related to the electronic bandwidths of the HD photodiodes and amplification-subtraction stages, and of any component on the CV acquisition path. In standard electronic design, $G_0$ and $f_c$ respect the constant product Gain-Bandwidth, a compromise is thus generally done to ensure fast detector response (high $f_c$) and an amplified signal with a high signal-to-noise ratio with respect to the electronic noise floor (easier with higher $G_0$). 

The higher is $f_c$, the better the detector will be able to retrieve information on fast signals encoded in high frequency components. In the case of heralded protocols, an optimal detection of the NG states requires the HD to be able to fully access information carried by the ideal detection mode. This means that the detector must be able to retrieve frequency components over the full bandwidth $B$ of $u_\text{id}(t)$~\cite{Gouzien_2018, kumar_versatile_2012}: the more $f_c$ exceeds $B$, the safer, in principle, is expected to be the HD measurement of the state. \\
Note that typical detection bandwidths in quantum optical experiments are on the order of a few hundred MHz (see, e.g.,~\cite{Takase:22, MorinHybrid2014}). In this work, we consider $f_c$ within this range, that is what is presently accessible to standard commercial detectors. At the same time, it is worth mentioning that nonlinear optical techniques can be used to enable broadband detection. Optical parametric amplifiers (OPAs) working as phase-sensitive amplifiers has been used to measure squeezed-light~\cite{Inoue2023} and NG states~\cite{Sonoyama2025} over bandwidths up to $\sim 40$~GHz. Very recently multimode OPAs have been used to detect broadband, multimode squeezing without requiring fast balanced detectors~\cite{CHEKHOVA2025}

\subsection{Digital Data sampling}
The analog signal that is available at the HD output is sent to the oscilloscope for the data acquisition. At each trigger event, a string of data is delivered, corresponding to the digital version, $v_i(t)$, of the HD output over a time duration $\Delta T$ (see Figure~\ref{fig:enter-label}-c). We can define the sampling rate frequency $f_s$ as the number of samples taken per second from the (continuous) HD output and then converted into a digital signal. A correct choice of $f_{s}$ is essential to prevent aliasing effects and distortions on the digitized photo-current~\cite{shannon_communication_1949, nyquist_certain_1928}. In practice, high sampling rate also guarantees a better robustness with respect to signal fluctuations due to residual electronic noise floor.

Since the HD acts as a filter, spectral components at its output are limited at most to frequencies of the order of $f_c$. Nyquist–Shannon theorem~\cite{shannon_communication_1949, nyquist_certain_1928} states that a temporal signal of spectral bandwidth $f_c$ can be faithfully reconstructed only if it is sampled at a frequency $f_s$ such that $2 f_{c} \le f_s$. The conditions for the reliable CV acquisition and data sampling of a detection mode with a bandwidth $B$, can thus be convenient summarized as: 
\begin{equation}
\label{condizioni}
  B\ll 2 f_c\leq f_s.
\end{equation}
As explained, $B$ is determined by the optical setup via the interplay between the ESS emission bandwidth and the optical filter, while $f_c$ and $f_s$ are determined by the electronic components on the CV acquisition chain. Eq.~\ref{condizioni} thus expresses in a simple way the parameters of interest to deal with temporal features; the CV acquisition chain must be designed so as to satisfy is as much as possible. Our scope is to understand what "as much as possible" means in the context of heralded state preparation and, for a given $u_\text{id}(t)$ (\emph{i.e.} $B$) what is the impact of $f_c$ and $f_s$ on the state reconstruction.

\section{Exploitation of CV measurement results}
The methodology usually used for full state reconstruction, and that we will also adopt here, is the following. The acquisition procedure detailed so far is replicated $N$ times. Provided the LO phase is scanned at an adequate speed~\cite{morin2012high, melalkia_plug-and-play_2022}, each digital trace $v_i(t)$ corresponds to one heralding event and refers to a different quadrature measurements at $\vartheta_i$ randomly taken in $[0,2\pi]$. The collection of $\{v_i(t)\}$ is exploited by means of digital data processing to gain information on $u_\text{id}(t)$ and, then, to use this information in quantum state tomography. In this section, we will discuss both the steps with reference to very common procedures that are reported in the literature~\cite{asavanant2017generation, morin2013experimentally, FurusawaGKP2024, MorinHybrid2014} and that we implemented in our experiments on NG state generation~\cite{melalkia_plug-and-play_2022}.

\subsection{Reconstruction of the detection mode}\label{detmoderec}

The ideal detection mode is a real function of time defined by the optical setup (ESS+filters) and normalized so that $\int u^2_\text{id}(t) dt = 1$~\cite{melalkia_plug-and-play_2022}. In realistic scenarios, theoretical estimations of $u_\text{id}(t)$ can differ from its actual shape due to imperfections or incomplete knowledge of the exact experimental parameters. In this cases, the actual profile of the detection mode can be inferred directly from homodyne data. In this kind of technique, quadrature measurements are post-processed and exploited in the numerical maximization of a given function~\cite{Molmer2007}. Here, we will follow a very popular technique~\cite{morin2013experimentally, yoshikawa2017purification, FurusawaGKP2024, melalkia_plug-and-play_2022} that relies on the optimization of the variances of the heralded state's quadratures and on a multimode analysis~\cite{morin2013experimentally}. HD data corresponding to different heralding events are used to compute the matrix $\langle v_i (t) v_i(t')\rangle$ of temporal-autocorrelation functions and subsequently to perform its eigenfunction expansion: the eigenfunction associated to the highest eigenvalue directly provides a reconstruction $u(t)$ of the optimal temporal profile. Reconstructed temporal functions are normalized to have $\int u^2(t) dt = 1$.  

This method can be applied to quantum state engineering experiments for which extracting the temporal mode is a central issue~\cite{morin2013experimentally, FurusawaGKP2024, melalkia_plug-and-play_2022}. The downside of it is that, evidently, imperfections in HD data acquisition or digitization reverberate on the temporal profile reconstructed by data~\cite{yoshikawa2017purification}. A critical point stands, thus, in assessing how, and with what consequences, $f_c$ and $f_s$ can lead to possible mismatches between the computed $u(t)$ and the mode $u_\text{id}(t)$ defined by the optical setup~\cite{asavanant2017generation, morin2013experimentally, melalkia_plug-and-play_2022}. 

\subsection{Digital signal processing for quantum state tomography}\label{tomoproc}
In CW regime, experimental data $\{v_i(t)\}$ contain, \emph{a priori}, information on all modes that are accessible to the HD detector, including those do not carry the NG state. The projection into the desired temporal mode is done in post-treatment done by digitally weighting each temporal trace $v_i(t)$ with the function, $u(t)$. Each tomographic point $X_i$ is thus obtained as:
\begin{equation}
     X_i = \int u(t)v_i(t)\mathrm{d}t.
     \label{integre}
\end{equation}  
This method allows obtaining $N$ tomographic points from the digital traces, each corresponding to a different heralding event and to the acquisition of a quadrature at a different $\vartheta_i$. The $\{X_i\}$ are organized by values of $\vartheta_i$ and used to perform the quantum tomography of the NG state density matrix and of its Wigner function. We use for the tomography a \textit{Maximum Likelihood} iterative algorithm , that is the most widespread (and easy to implement) one in recent experiments~\cite{Lvovsky_Tomo2009}.
Interestingly, reducing $f_c$ and $f_s$ affects the matching of $u(t)$ with $u_\text{id}(t)$ as well as the quality of $v_i(t)$, that can be deformed and temporally smoothened. Both effects thus lead to a degradation of tomographic points $X_i$ that eventually reflects on the quality of reconstructed NG states. 

\section{Methods: experimental data acquisition and pre-processing }

To test the impact of the temporal parameters of the CV acquisition we used the dataset $\{v^\text{exp}_i(t)\}$ that we measured in an experiment on the heralded generation of Schroedinger kitten states via single photon subtraction from a squeezed vacuum state. Details on the setup are in Ref.~\cite{melalkia_plug-and-play_2022}. Strategically, this architecture implements an key method to generate NG state, whose idea has been applied to schemes of increasing complexity~\cite{ZavattaHybrid2014, MorinHybrid2014, FurusawaGKP2024, lvovsky2020production}. Using experimental data instead that simulated ones allows our analysis to have solid basis in real-world systems. In addition, Schroedinger kitten states possess phase-sensitive Wigner functions. This allows us to explore the situation of NG features that depend explicitly on the optical phase and therefore provides a more stringent and informative test of our analysis tools. At the same time, the methods presented in this work apply in a general way to multiple experimental schemes, as both the technique for the reconstruction of $u(t)$ and quantum tomography are currently used for a general class of NG state.

Data in~\cite{melalkia_plug-and-play_2022} are acquired with optimized parameters. The bandwidth of the optical cavity on the heralding path has been set to $\Gamma=9.3$\,MHz, and experimental data are acquired by a HD with $f^{\text{exp}}_c=301$~MHz and sampled by a fast oscilloscope at a rate $f_s^{\text{exp}}=5$~Gsps well above the Shannon-Nyquist theorem. This choice of values makes it possible to obtain from experimental HD data a reconstructed profile that fully matches the $u_\text{id}(t)$ that was theoretically expected for our optical setup.
The signal-to-noise ratio of the CV detection was 12\,dB above the electronic noise. A total of $N=$\,43000 signal traces of duration 5\,$\mu s$ have been experimentally acquired and processed to obtain the quadrature measurements required for quantum tomography. 
The time $t_0$ appearing in the asymmetric profile of $u_\text{id}(t)$ in Fig.~\ref{fig:enter-label}-b represents the delay between the time at which the heralding signal is delivered and the instant at which the trigger of the CV acquisition chain fires. This is due to optical and electrical delays that pile up between the heralding and heralded paths and it is independent on $f_c$ and $f_s$. Its value, experimentally measured once can be kept constant for all data treatment~\cite{melalkia_plug-and-play_2022}. For each trigger signal, homodyne traces are acquired over times longer than the expected temporal extent of the $u_\text{id}(t)$ (of the order of $\approx \Gamma ^{-1}$): we select out of this long acquisition, a signal $v^\text{exp}_i(t)$ of duration $\Delta t$, that we use to reconstruct the temporal profile $u(t)$ and in Eq.~\ref{integre}. Data outside this slot contain information about squeezed vacuum: they are used to associate to each $v^\text{exp}_i(t)$ the correct LO phase $\vartheta_i$ and estimate the phase corresponding to each acquired quadrature data.
The duration of $\Delta T$ must be taken above $\Gamma ^{-1}$ to be sure of capturing the full temporal mode and avoid loosing part of the information on the NG state. However, $\Delta T $ unnecessarily long would introduce on data additional noise due to residual fluctuations outside the temporal region of interest. In our case, only data $\{v^\text{exp}_i(t)\}$ within the time window $\approx[-1.25\,\mu$s,$-1\,\mu$s] (corresponding to a $\Delta T=0.25\,\mu$s) are post-processed to extract the heralded NG state quadratures. The raw quadrature data are corrected beforehand to take into account the overall detection efficiency of the homodyne detection $\eta_{HD}=0.72$~\cite{melalkia_plug-and-play_2022, lvovsky2020production}.

In the following, experimental $\{v^\text{exp}_i(t)\}$ is digitally post-treated to push $f_c$ and $f_s$ toward lower values and to assess the resilience of the entire digital signal process to a degraded CV detection. The procedure is articulated in two steps: 
\begin{enumerate}
\item[-] Firstly, we simulate the effect of detector bandwidth $f_c\leq f^\text{exp}_c$ by adding a numerical filter on each $v^\text{exp}_i(t)$.
This digital filter has the same shape of the Butterworth function of \ref{TransfertF} and it is applied numerically using the Python library Scipy. This will lead to $v_i(t)\approx FT^{-1}[H(f,f_c)\cdot \Tilde{v}^\text{exp}_i(f)]$, where $\Tilde{v}^\text{exp}_i(f)=FT[\Tilde{v}^\text{exp}_i(t)]$ is the Fourier Transform of the temporal traces acquired in the experiment with optimal detection parameters.
\item[-] Secondly, we simulate a low sampling rate $f_s\leq f^\text{exp}_s$, by manually under-sampling the $v_i(t)$ by extracting from it one point out of $n$. Note that this processing gives access to the specific sampling frequencies $f^{(n)}_s=f^\text{exp}_s/n$. To reproduce the aleatory position of the trigger within the temporal trace and avoid systematic effects, we randomize the position of the first suppressed point from one experimental trace to the other.
\end{enumerate}

The dataset obtained by applying these procedure undergoes the digital data treatement explained in the previous paragraph. In this way, we show the effect of both $f_c$ and $f_s$ on the reconstructed $u(t)$ and, subsequently, on the quality of states obtained from the quantum tomography.
As a final remark, note that in heralded protocols, filtering stages (electronic or digital) are routinely used to define the temporal mode of the detected signal: applying a digital filter on a classical photocurrent is equivalent to using an analog filter with the same transfer function, yielding identical mode shaping. Consequently, modeling the finite detector bandwidth with a well-defined digital filter—such as the second-order Butterworth low-pass filter used here—is fully consistent with standard experimental practice.

\section{Effect of the variation of $f_c$ and $f_s$}

\subsection{Reconstruction of the detection mode $u(t)$}
We start by studying the influence of $f_c$ and $f_s$ on the reconstruction of the detection temporal mode. To do so, we apply the procedure described in subsection~\ref{detmoderec} to pre-treated data. We compare the resulting $u(t)$ to the optimal one that matches $u_\text{id}(t)$ and is obtained for native experimental values. For simplicity, we will discuss our results in terms of the optical filter bandwidth $\Gamma$, that for the experimentalists is of more direct access than $B$. Note that, as in general $\Gamma<B$, deformation effects start becoming visible on the $u(t)$ when $f_c$ and $f_s$ approach $B$, that happens at few multiples of $\Gamma$.\\

\begin{figure}[ht]
\centering
\includegraphics[width=\columnwidth]{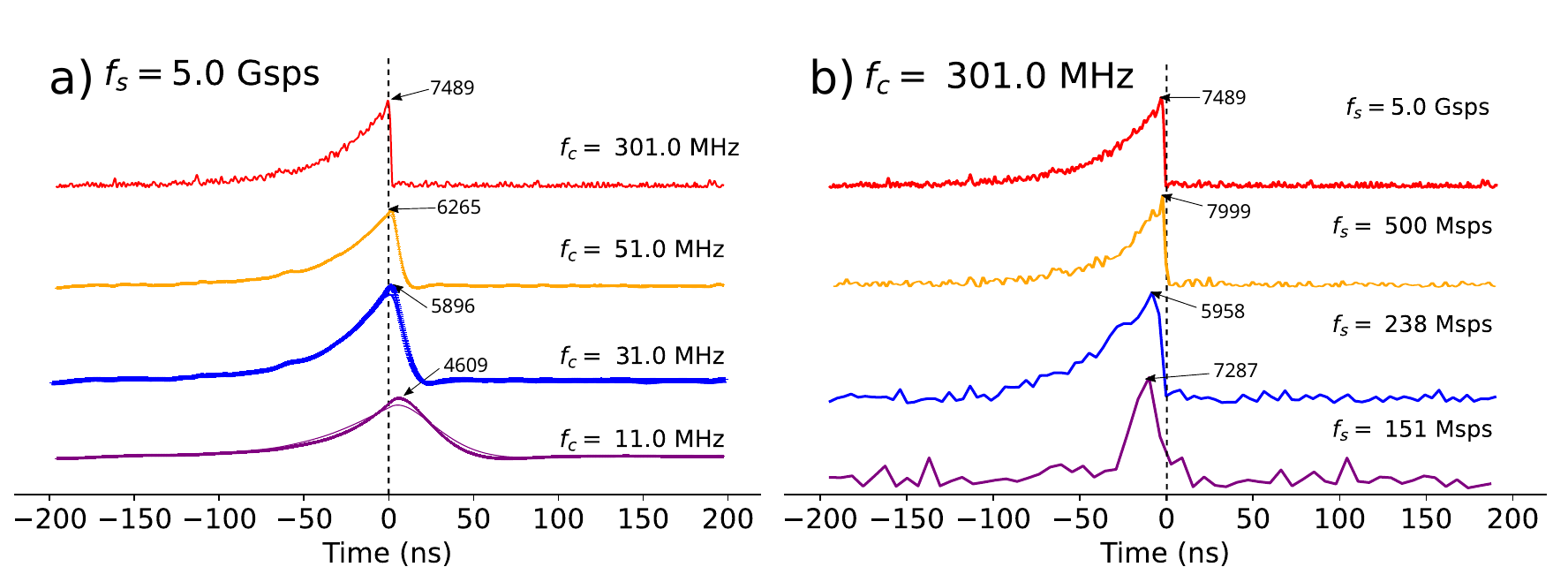}
\caption{Temporal modes $u(t)$ in a.u. extracted from experimental data post-treated to simulate the effect of different HD cutoff frequency (a) and sampling rates (b). All profiles are normalised to have $\int u(t)^2dt=1$. The peak height is indicated for each $u(t)$. The profile corresponding to $f_c=301$\,MHz and $f_s=5$ Gsps (in red) is the one obtained with native experimental data and it fully matches the analytically computed $u_\text{id}(t)$~\cite{melalkia_plug-and-play_2022}. The thinner lines in figure (a) show the temporal profiles that are simulated by applying the Butterworth filter to the ideal detection mode, \textit{i.e.} $u_\text{fc}(t)= FT^{-1}[H(f,f_c)\cdot \tilde{u}_\text{id}(f)]$.}
\label{TMrecostructed}
\end{figure}

\textbf{Limited HD bandwidth}. To illustrate the influence of $f_c$ on $u(t)$, data are pretreated to reproduce the effect of an HD with a cutoff frequency between $11$\,MHz and $291$\,MHz. The resulting $u(t)$ are shown in Fig.~\ref{TMrecostructed}-a. The sampling rate is kept fixed at its experimental value $f_s^{\text{exp}}=5$\,Gsps. For each of the reported curves, we computed the mismatch, of the reconstructed profile $u(t)$ with the one corresponding to $f_c^{exp}$ and $f_s^{exp}$ that well approximated the ideal one $u_\text{id}(t)$. This is computed as $1-\eta$ with $ \eta \propto \int u_{id}(t)u(t)\mathrm{d}t$.

Slow HDs with low $f_c$ are not able to follow the fast dynamics of data relative to $u_\text{id}(t)$. Accordingly, when reducing $f_c$, a more and more distorted version of $u_\text{id}(t)$ is delivered by the reconstruction algorithm. Our analysis shows that for our set of data, the reconstruction is robust as long as $f_c\gtrsim 5 \Gamma$ (\textit{i.e.} $f_c$>50\,MHz) with a mismatch of the reconstructed profile with the optimal one staying below few percents. For $f_c$ below this value, the sharp edge of $u_\text{id}(t)$ is progressively smoothened, eventually leading to a flattened and almost symmetric $u(t)$ when $f_c\approx \Gamma$. For our data, this corresponds to the profile $u(t)$ at $f_c=11$\,MHz, whose width is almost the double of the one of the ideal profile corresponding to a mismatch with the initial experimental profile above $30\%$. Such a behavior is well explained by Eq.~\ref{TransfertF}. Steep temporal signals correspond, in the Fourier transform domain, to broad spectral functions containing high frequency components. As the HD filters out fast frequency components at $f>f_c$, the effect of low $f_c$ is particularly evident on the sharp edge of the asymmetric profile that is encoded on higher frequency components.\bigskip

The dependency of $u(t)$ on $f_c$ is due to the effect of the HD temporal response on the data fed to the reconstruction algorithm. Remarkably, $u(t)$ corresponding to different $f_c$ can be estimated by merging our theoretical knowledge on both the optical and electronics setup. The thinner lines in Fig.~\ref{TMrecostructed}-a) show the temporal profiles $u_\text{fc}(t)$ that are obtained by applying the Butterworth filter $H(f,f_c)$ to the theoretical detection mode, \textit{i.e.} $u_\text{fc}(t)= FT^{-1}[H(f,f_c)\cdot \tilde{u}_\text{id}(f)]$, where $\tilde{u}_\text{id}(f)$ is the Fourier transform of $u_\text{id}(t)$. As it can be seen, $u_\text{fc}(t)\approx u(t)$ for all $f_c$, with an almost perfect matching for high $f_c$ when the effect of $H(f,f_c)$ is negligible.\bigskip

\textbf{Limited sampling rate}. To study the effect of the sampling rate on $u(t)$, we consider pre-treated data with a fixed $f^\text{exp}_c=301$\,MHz and with $f_s$ going from 2.5\,Gsps to 151\,Msps. Note that $f_s$ of few hundreds of Msps correspond to current standard performances of mid-range oscilloscopes; low $f_s$ have been included in the analysis only to illustrate the effect of very poor sampling rates.

Figure~\ref{TMrecostructed}-b) reports obtained results. Simulations show that the reconstructed $u(t)$, are quite robust against the reduction of the sampling rate, with no relevant changes even when $f_s$ goes down to $555$~Msps, that is almost one order of magnitude smaller than the original value $f^\text{exp}_s$. Artifacts and increasing deformations appear on $u(t)$ for $f_s\leq500$~Msps: this reflects the progressive violation of the condition $2f_c\leq f_s$ expressed in Eq.~\ref{condizioni}. 
Remarkably, the effect of $f_s$ on the reconstruction of the detection mode profoundly differs from the one of $f_c$; $u(t)$ tends to keep its asymmetry and original width, and the reduction of $f_s$ essentially only makes the profile noisier. At very low $f_s$, major under-sampling effects combine with such noises and make it difficult to obtain an accurate reconstruction of $u(t)$, eventually leading to scattered profiles as for $f_s\approx 151$\,Msps. \bigskip

\subsection{Tomographic reconstruction}
For each choice of $f_c$ and $f_s$, we obtain a set of $N$ = 43000 tomographic points: each of them is computed by using a pre-treated $v_i(t)$ and a reconstructed $u(t)$ in the integral of Eq.~\ref{integre}. Tomographic points undergo a maximum likelihood algorithm with 2000 iterations. This leads to the reconstruction of the Wigner function that is actually accessible to a CV detection chain with limited temporal performances~\cite{Lvovsky_Tomo2009}. 
Original data $\{v^\text{exp}_i(t)\}$ refer to the generation of a Schrodinger kitten state, whose Wigner function is negative at the origin. The negativity of the Wigner function is, indeed, one of the most striking non-classical aspects of quantum states and it is a necessary feature for quantum speedup in continuous variable quantum information~\cite{Chabaud_2021}. Compared to other possibly witnesses, as the fidelity with the target state, the Wigner negativity also has the advantage of being particularly sensitive to losses and imperfections and it avoids ambiguous interpretations of data that would be associated, for instance, to an analysis based on state purity~\cite{Kenfack2004}. 

We start by focusing on it as a parameter to assess the robustness of the tomographic reconstruction against CV acquisition chain limitations. We implemented the Maklik algorythm on tomographic data corresponding to different choices of $f_c$ and $f_s$. In our analysis, the number of available tomographic points and the maximum dimension of the targeted density matrix are independent on $f_c$ and $f_s$ and were fixed from the original set of experimental data. We observed minor variations on the number of iterations required to the algorithm to converge (typically $\approx$\,200), but no systematic effect with $f_c$ or $f_s$: for all the Wigner reconstructions, we set to number of iterations (1000) well above the value required for the convergence.\bigskip

\begin{figure}[htpb]
\centering
\includegraphics[width=\columnwidth]{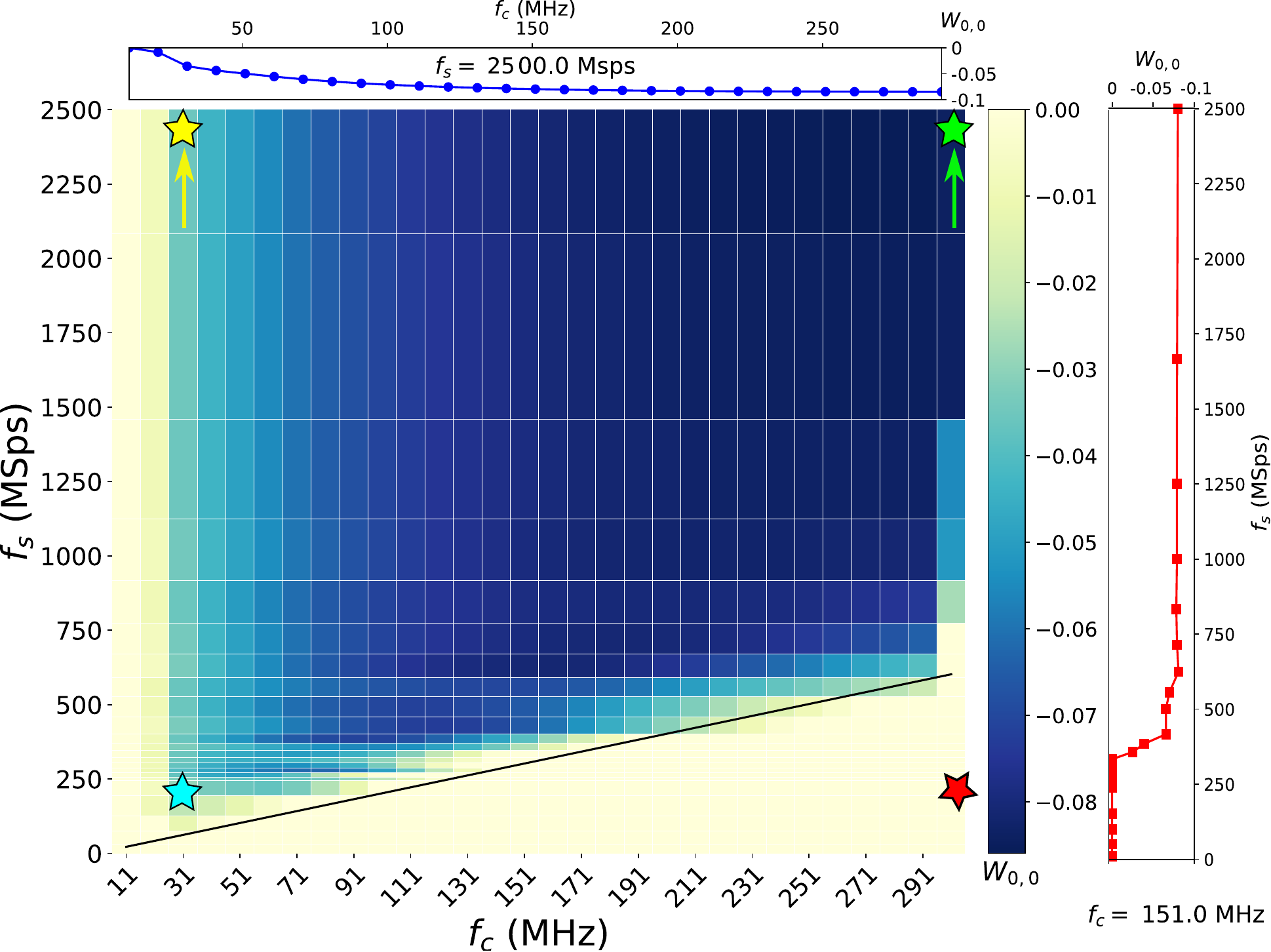}% Here is how to import EPS art
\caption{Negativity of the Wigner function at the origin as a function of the CV detector bandwidth ($f_c$) and of the sampling rate ($f_s$). The values of $f_s$ have been limited to values below 2.5\,Gsps as no relevant change is observed in $W_{0,0}$ between $f_s$= 2.5\,Gsps and 5 \,Gsps.} Lateral inset: degradation of $W_{0,0}$ with decreasing $f_s$ for a fixed $f_c= 151$\,MHz.  Top inset: evolution of $W_{0,0}$ with $f_c$ for $f_s= 2.5$\,Gsps. The stars in the figure indicated the points corresponding the Wigner functions of Fig.~\ref{fig:W}-a (green), -b (red star), -c (yellow star) and -d (cyan star). The arrows close to the green and yellow star indicate that the curves in Fig.~\ref{fig:W}-a and -c rather refers to $f_s$= 5\,Gsps, that is out of the scale of this figure.\label{fig:neg} 
\end{figure}

Fig.~\ref{fig:neg} shows the value at the origin of the Wigner functions referring to sampling frequencies $f_s$ from 10~Msps to 5~Gsps and HD bandwidths $f_c$ from 11~MHz to 301~MHz. The starting point is the optimal $W^\text{exp}_{0,0}=-0.084\pm0.004$, experimentally obtained with dataset corresponding to $f^\text{exp}_c$ and $f^\text{exp}_s$. For any fixed $f_c$, a rapid transition from negative to positive Wigner functions takes place when $f_s$ does not respect the Shannon-Nyquist condition: the border at $2 f_c\leq f_s$ is indicated by the black line in the figure. In a reciprocal way, when $f_s$ is very low, forcing $f_c$ above the Shannon-Nyquist criterion only introduces high frequency noise contribution and aliasing that deform the NG Wigner function. An example of the degradation of $W_\text{0,0}$ with decreasing $f_s$ is provided by the lateral inset of Fig.~\ref{fig:neg} for a fixed $f_c= 151$\,MHz: the Wigner negativity abruptly goes from -0.08 to -0.04 at $f_s\approx 300$\,MHz and eventually to 0 for $f_s\leq 200$\,MHz. In this regard, note that, beside introducing noise in the reconstruction of $u(t)$ (see Fig.~\ref{TMrecostructed}-b), low $f_s$ reduce the number of points with whom $v_i(t)$ is sampled and, in turn, the number of points usable for the (digital) integration of Eq.~\ref{integre}. As a matter-of-fact, even if the total number of tomographic points stays the same (N =\,43000), the tomographic reconstruction is affect by reduced precision in the way each point $X_i$ is obtained.  \bigskip

\begin{figure}[ht]
\centering
\includegraphics[width=\columnwidth]{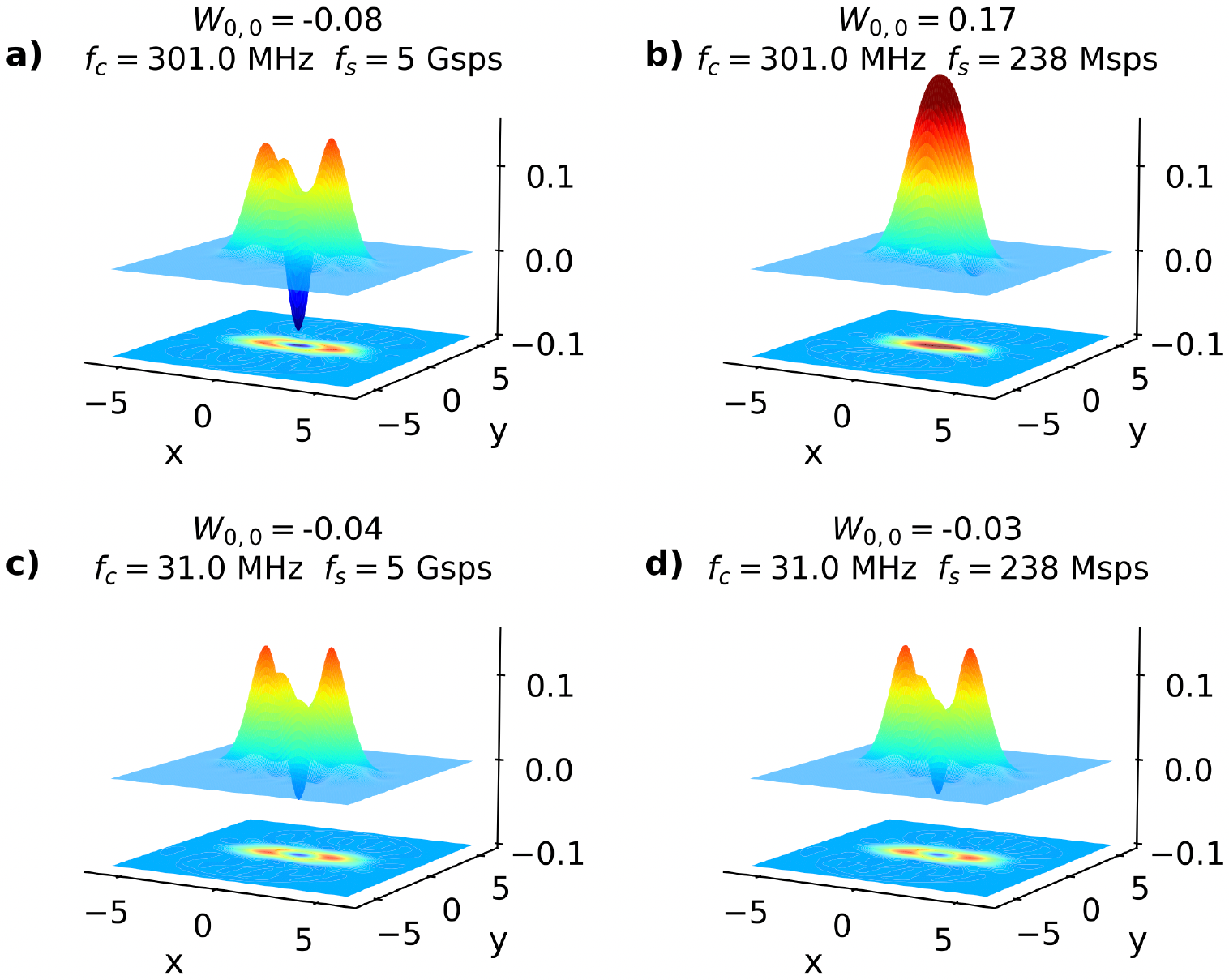}% Here is how to import EPS art
\caption{Reconstructed Wigner functions $W(x,y)$ 
of the heralded NG state 
corrected by the homodyne detection efficiency $\eta_{HD}=0.72$. 
(a) Optimal state at $f_c=301$\,MHz and $f_s$=5\,Gsps (untreated experimental data). 
(b) Reconstructed state at $f_c=301$\,MHz and $f_s=238$\,Msps.
(c) Reconstructed state at $f_c=31$\,MHz and $f_s=5$\,Gsps.
(d) Reconstructed state at $f_c=31$\,MHz and $f_s=238$\,Msps.}\label{fig:W}
\end{figure}
In the zone where Shannon-Nyquist theorem is satisfied, Fig.~\ref{fig:neg} shows that, as a general rule, the state is strongly affected by the information loss due to the filter effect of $H(f,f_c)$. When reducing $f_c$, the HD filters high frequency components of electronics signals and smoothens both $v_i(t)$ and $u(t)$ loosing some information on temporal features of the NG state: such an effect induces a degradation of the reconstructed state. This is witnessed by the progressive reduction of negativity associated to low $f_c$ that can be seen for any $f_s$. Nevertheless, the Wigner negativity exhibits a certain robustness against the HD detection bandwidth. Interestingly, even when the reconstruction of the asymmetric profile is distorted by a low $f_c$, if the HD is able to reconstruct at least the slow side of $u_\text{id}(t)$, the Wigner function stays Non-Gaussian and, to a lesser extent, negative. To give a concrete example, we can discuss this for $f_s=$5\,Gsps, \textit{i.e.} for the sampling rate used in our original experiment~\cite{melalkia_plug-and-play_2022}. As seen in the previous paragraph, for our experiment, the condition for a non distorted $u(t)$ is $f_c\gtrsim 5\Gamma$ \textit{i.e.} $f_c$>50\,MHz (see Fig.~\ref{TMrecostructed}-a). The top inset of Fig.~\ref{fig:neg} shows the evolution of $W_\text{0,0}$ with $f_c$: the negativity is reduced by a factor $\approx$2 when the bandwidth of the detector passes from 301\,MHz to 31\,MHz (\textit{i.e} it is decreased by a factor 10). At the same time, degraded but still negative NG Wigner functions are obtained even for detection bandwidths lower than $50$~MHz. Similar results are found at lower $f_s$ above the black line depicted in Fig.~\ref{fig:neg}. 
We note that a behavior similar to the one reported in Fig.~\ref{fig:neg} is observed for the fidelity between the Wigner obtained for optimal $f_c$ and $f_s$ and those corresponding to lower quality data: a fidelity above 80$\%$ is preserved if the Shannon Nyquist criterium is satisfied, and, for the choices of $f_c$ and $f_s$ satisfying this condition, a progressive decrease of the fidelity is observed when the bandwidth $f_c$ is reduced.\bigskip

Results on the $W_\text{0,0}$ are confirmed and enriched by Fig.~\ref{fig:W}. This compares the Wigner functions corresponding to different $f_s$ and $f_c$ with the one in Fig.~\ref{fig:W}-a), corresponding to the original experimental dataset $\{v^\text{exp}_i(t)\}$~\cite{melalkia_plug-and-play_2022}. Fig.~\ref{fig:W}-b) shows in detail the dramatic effect of a poor sampling rate for $f_c= f^\text{exp}_c$ and $f_s=238$\,Msps$<2f_c$. Although the Shannon--Nyquist criterion is well known in digital data processing, its impact on Wigner function reconstruction and quantum tomography has never been investigated. The reconstructed Wigner of Fig.~\ref{fig:W}-b is totally positive and strongly deformed with respect to the initial one, showing that the information on the generated state is completely lost. To have an intuition of this behavior, we recall that data on the non-Gaussian state are retrieved by reweighting, via the $u(t)$, a broader dataset that mostly contains information about the initial squeezed state to which the photon subtraction is applied. In our experiment, squeezing is generated in a single-pass configuration over an optical bandwidth of teraHerz, with constant squeezing levels over relevant bandwidths (essentially given by $f_c$). The noise of the measured squeezed vacuum states can therefore be considered as a white, with different spectral components carrying the same squeezing information. This makes the squeezing measurement robust to the loss of high-frequency information associated to the violation of Nyquist-Shannon criterium as well as to the specific profile of $u(t)$ used in Eq.\,\ref{integre}. Conversely, the Non-Gaussian state only exists in a mode determined by the temporal profile $u(t)$ whose spectral information is not flat, and critically depends on high-frequency signatures introduced by photon subtraction. In this case, low $f_s$ induce a bad mode selection, due to the deformation of $u(t)$, but also a critical suppression of relevant spectral information on the $v_i(t)$ themselves. In short, due to the under-sampling, the dominating gaussian squeezing information remains largely preserved in the state reconstruction, while the more delicate features associated with the photon-subtracted state are degraded and eventually disappears, leading to a positive Wigner. Figs.~\ref{fig:W}-c) and -d) illustrate the effect of $f_c$. Reported experimental works generally take the precaution of working with a HD bandwidth way higher than the optical filter width ($\Gamma\ll f_c$)~\cite{melalkia_plug-and-play_2022}. Our analysis shows that, when $f_c$ approaches $\Gamma$, it is still possible to observe strongly non-Gaussian Wigner function exhibiting some negativity. This behavior is confirmed by the Fig.~\ref{fig:W}-c) and -d) for data acquired with $f_c=$31\,MHz and with $2 f_c<f_s$ (respectively $f^\text{exp}_s$ for fig-c) and $f_s=238$\,Msps for fig-d) ). When $f_c$=31\,MHz, Wigner functions negativity halves with respect to the optimal value obtained for $W^\text{exp}_{0,0}$. Nevertheless, the reconstructed functions clearly exhibit a non-Gaussian behavior and preserve the two lobes that are characteristic of the target Schroedinger cat state. Minor differences are observed for the two considered values of $f_s$, indicating that the effect on the Wigner is mostly due to $f_c$. These results indicate that, for a given temporal mode $u_\text{id}(t)$, \textit{i.e.} for a given $\Gamma$, under certain conditions, reliable NG state reconstructions can be achieved using slower homodyne detectors than commonly assumed. This finding has direct practical relevance for experimental implementations, where bandwidth limitations can constitute a critical constraint. Reciprocally, for a fixed $f_c$, constraints on narrow $\Gamma$ can be relaxed, thus allowing experiments to work with broader optical filters on the heralding path and improving the overall heralding rate of the experiment. At the same time, we remark that satisfying the Shannon-Nyquist condition on the sampling rate is of paramount importance, as its violation can lead to substantial distortions in the reconstructed quantum states. In this regard, it is worth comparing fig-b and fig.-d: for a given sampling rate $f_s$=238\,Msps, degrading $f_c$ so as to make the acquisition satisfy condition of equation~\ref{condizioni}, allows the Wigner function to recover its negativity, albeit in a degraded form due to the reduced \(f_c\) compared to the optimal experimental value. The predominance of the Shannon--Nyquist criterion over the degradation of \(f_c\) is clearly visible also in Fig.~\ref{fig:neg} and was never discussed so far.\\

 As a final remark, the comparison between the results for $u(t)$ and the Wigner functions shown in Figs.~\ref{TMrecostructed} and \ref{fig:W} suggests that an inaccurate reconstruction of the mode $u_{\text{id}}(t)$ does not necessarily lead to a significant degradation of the reconstructed quantum state. To investigate this point, we replaced $u(t)$ with the theoretical profile $u_{\text{id}}(t)$ in Eq.~\ref{integre}, while processing the data $v_i(t)$ so as to vary $f_c$ and $f_s$. We found that, for all values of $f_c$ and $f_s$, substituting the experimentally reconstructed mode with its ideal counterpart leads to only marginal improvements in the Wigner function. This indicates that the observed degradation is more strongly linked to the impact of the CV detection chain on $v_i(t)$ than to inaccuracies in the mode-profile estimation. Consequently, when the ideal temporal mode can be reliably determined from theory or previous experiments, it may be safely employed in Eq.~\ref{integre} for tomographic reconstruction. A more quantitative assessment of this conclusion would require a dedicated theoretical model, which lies beyond the scope of the present work and will be addressed in a future theoretical study.
\begin{comment}
\begin{figure}[ht]
\centering
\includegraphics[scale=0.4]{comparaison125.png}% Here is how to import EPS art
\caption{\label{fig:Wcompar} Reconstructed Wigner functions $W_{NG}(x,p)$ with $u(t)$ and with $u_\text{id}(t)$}
\end{figure}
\end{comment}
\section{Conclusion}
We investigated the impact of limited temporal performances of the CV acquisition chain on the detection of heralded NG states in CW regime. We focused in particular on how electronic bandwidth and digital sampling rate affect the way NG states are detected, treated and eventually reconstructed by quantum tomography. Simulating digital data treatments  with varying parameters allowed us to identify, for a fixed heralding bandwidth, $\Gamma$, the conditions on $f_c$ and $f_s$ to respect to retrieve NG and negative quantum states out of the entire detection process. Our analysis shows that, provided Shannon-Nyquist criterium is respected, NG features are preserved even when the detection bandwidth is closer to the optical bandwidth of filters on the heralding path. The state negativity is however degraded by the loss of information associated to slow temporal responses. By assessing and quantifying these effects, our analysis allows to understand results that can be achieved by exploiting standard devices with limited performance and paves the way to a more efficient and aware use of experimental resources. Conceptually, our analysis has allowed to show the strengths and criticalities of different working regimes and highlighted behaviors that were not initially apparent. By doing so, we open a new direction of investigation and show that meaningful quantum-state reconstruction can remain possible under significantly relaxed experimental constraints.

\section{Funding}
Authors acknowledge financial support from European Union by means of Fond Européen de développement regional (FEDER, project OPTIMAL), the Plan France 2030 through the project ANR-22-PETQ-0013 and Université Côte d'Azur via CSI and Academie RISE funding.

\section{Acknowledgments}
Virginia D'Auria acknowledges the Institut Universitaire de France (IUF). Mohamed Faouzi Melalkia is currently at Laser Systems Laboratory EMP, Algiers, Algeria.

\section{Disclosures}
The authors declare no conflicts of interest.

\section{Data availability} 
Data underlying the results presented in this paper are not publicly available at this time but may be obtained from the authors upon reasonable request.

%%%%%%%%%% If using BibTeX:
%
%
\end{document}